\documentclass[sn-nature]{sn-jnl}
\usepackage{amssymb}
\newcommand\authormark[1]{\textsuperscript{#1}}

\usepackage{graphicx}%
\usepackage{multirow}%
\usepackage{amsmath,amssymb,amsfonts}%
\usepackage{amsthm}%
\usepackage{mathrsfs}%
\usepackage[title]{appendix}%
\usepackage{xcolor}%
\usepackage{textcomp}%
\usepackage{manyfoot}%
\usepackage{booktabs}%
\usepackage{algorithm}%
\usepackage{algorithmicx}%
\usepackage{algpseudocode}%
\usepackage{listings}%
\begin{document}

\title{Unified, Verifiable Neural Simulators for Electromagnetic Wave Inverse Problems}


\affil[1,2,3]{\orgdiv{Department of Electrical Engineering and Computer Sciences}, \orgname{University of California, Berkeley}, \orgaddress{\city{Berkeley}, \postcode{94720}, \state{CA}, \country{USA}}}

\author*[1]{Charles Dove}
\email{\authormark{1}charles\_dove@berkeley.edu} 
\author[2]{Jatearoon Boondicharern}
\email{\authormark{2}jboondicharern@berkeley.edu}
\author*[3]{Laura Waller}
\email{\authormark{3}waller@berkeley.edu}

\abstract{
Simulators based on neural networks offer a path to orders-of-magnitude faster electromagnetic wave simulations. Existing models, however, only address narrowly tailored classes of problems and only scale to systems of a few dozen degrees of freedom (DoFs). Here, we demonstrate a single, unified model capable of addressing scattering simulations with thousands of DoFs, of any wavelength, any illumination wavefront, and freeform materials, within broad configurable bounds. Based on an attentional multi-conditioning strategy, our method also allows non-recurrent supervision on and prediction of intermediate physical states, which provides improved generalization with no additional data-generation cost. Using this $O(1)$ time intermediate prediction capability, we propose and prove a rigorous, efficiently computable upper bound on prediction error, allowing accuracy guarantees at inference time for all predictions. After training solely on randomized systems, we demonstrate the unified model across a suite of challenging multi-disciplinary inverse problems, finding strong efficacy and speed improvements up to 96\% for problems in optical tomography, beam shaping through volumetric random media, and freeform photonic inverse design, with no problem-specific training. Our findings demonstrate a path to universal, verifiably accurate neural surrogates for existing scattering simulators, and our conditioning and training methods are directly applicable to any PDE admitting a time-domain iterative solver.
}
\maketitle
\vspace{-30pt}
\section{Main}
Electromagnetic (EM) wave simulators based on Maxwell's equations, known as full-wave simulators, provide a detailed quantitative understanding of light-matter interactions. Understanding and shaping these interactions is fundamental to various fields, from computational microscopy~\cite{Liu2017,Fang2009} and deep tissue imaging\cite{Ntziachristos2010,Qin2022,Zhao2022} to photonics~\cite{Lalau2013,Molesky2018,Yu2014} and metamaterials. Traditional full-wave simulation methods like the finite difference time domain (FDTD)~\cite{Yee1966} method are accurate but slow and computationally intensive, with compute requirements scaling directly with a system's spatial size. This is especially limiting for iterative inverse problems, such as image reconstruction\cite{Fang2009} and photonic inverse design\cite{Su2018,Molesky2018}, which generally require hundreds or thousands of consecutive, differentiated simulations. Solution of inverse problems of modern interest thus commonly takes hours or days, and in many situations, such as metamaterial design\cite{Yu2014,Kuznetsov2024,McClung2020}, full-scale solution is so time-demanding as to be practically impossible. Additionally, many imaging and sensing applications are time-sensitive. In such situations, simplified scattering models must be used \cite{Devaney1981,Chen2020} which trade physical accuracy for simulation speed. For these reasons, there is a pressing need for EM wave simulators which are fast but retain the quality of conventional full-wave methods.

Neural network (NN)-based simulators, also called neural simulators, have the ability to provide fast EM field predictions for a given source and refractive index (RI) configuration. Existing approaches\cite{An2019,Jiang2019,Yao2019,Nadell2019,Hegde2020,So2020,Inampudi2018,Martin2020,Liu2018,Gao2019,Jiang2020,Lim2022,Chen2022}, however, have three key limitations. Firstly, they are task-specific, with a given model addressing a narrowly defined class of systems, wavelengths, and illumination conditions. As such, any sufficiently distinct task necessitates its own distinct training process. For many applications, especially design tasks in which only a few designs are required, the computational cost and programming effort of this can easily outweigh the benefits of acceleration, and practitioners opt for slow but simple conventional simulators. 

Secondly, existing approaches have only been scaled to problems with modest numbers of degrees of freedom (DoFs), up to approximately 20\cite{Liu2018}. For this reason, these approaches have found the most success in structured design tasks, like grating coupler design\cite{Chen2022} and lens design \cite{Lim2022}, in which each DoF represents a bulk device parameter, such as the diameter of a nano-strip. Such approaches, however, fail to address freeform tasks, crucial for emerging photonic and metamaterial applications, in which each DoF represents the RI at a single voxel location and thousands of DoFs is the norm. Similarly, these approaches fail to address common reconstructive imaging tasks, such as reconstructive microscopy of biological tissue, for which materials lack predefined structure. Scaling existing techniques to higher numbers of DoFs while retaining satisfactory accuracy on relevant problems has been predicted to be remarkably challenging, with existing approaches exhibiting exponential growth in the size of the training dataset with DoF number\cite{Jiang2020}.

Finally, existing approaches have no rigorous mechanism for efficiently validating the accuracy of predictions. While physics-informed estimates based on residuals exist\cite{Chen2022,Lim2022}, these lack a quantitative correspondence with the true error. 
\begin{figure}
    \centering
    \includegraphics[width=\textwidth]{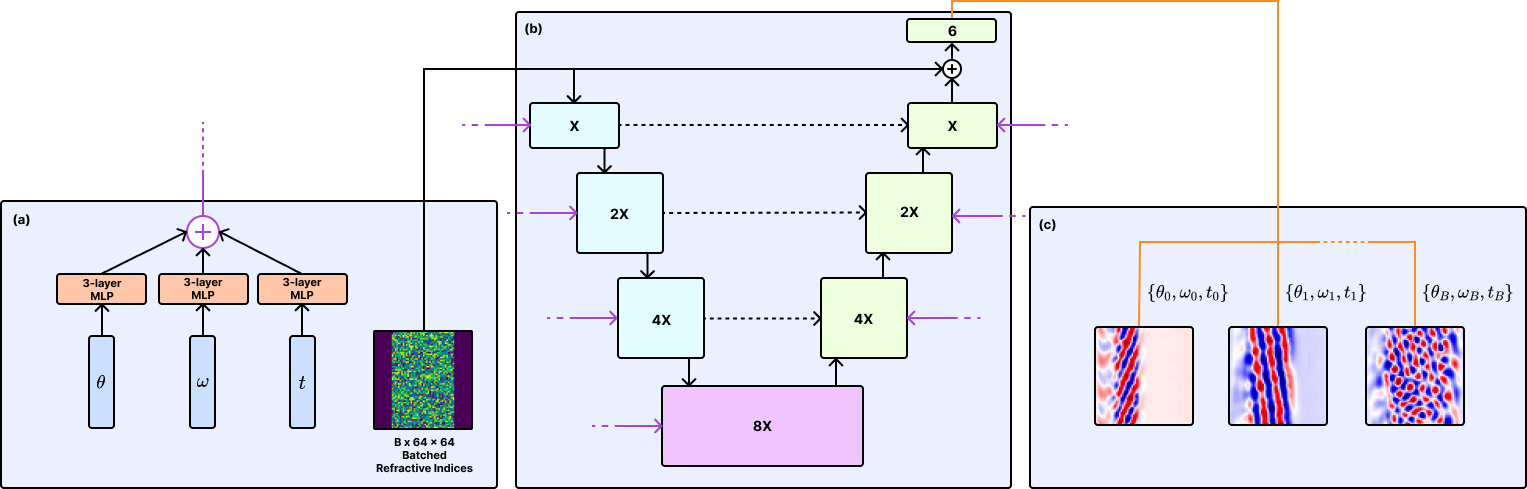}
    \caption{
Diagram of the UCMax model and conditioning approach. (a) Angle, wavelength, and timestep information are encoded, then fed into the model along with refractive index information. The orange boxes indicate a sinusoidal encoding layer followed by an MLP. (b) Model architecture. Each labeled square denotes a convolutional residual block followed by an attentional computation, with channel dimensions equal to a multiple of hyperparameter $X$, here chosen as 64. Light blue boxes incorporate a downsampling operation, green incorporate upsampling, and pink incorporate neither. (c) Predictions for all inputs are computed in a single batch, with no recurrence.}
    \label{fig:enter-label}
\end{figure}

In this paper, we present unified conditioning for Maxwell scattering (UCMax), a new, unified approach to the physics-guided neural simulation of EM wave scattering in freeform materials. Trained as a surrogate for FDTD, UCMax is capable of addressing arbitrary scattering simulations with thousands of DoFs, of any wavelength, any illumination wavefront, and voxel-based, freeform materials, each within broad bounds similar to those on the underlying FDTD solver. UCMax is based on an attentional multi-conditioning approach and is, to our knowledge, the first non-recurrent method to allow supervision on the intermediate physical states of the underlying physics solver, which we find provides significant generalization benefits for no additional data-generation cost.  Further, we use this intermediate prediction capability, which is computable in $O(1)$ time for any timestep, to derive a provable inference-time upper bound on prediction error, itself efficiently computable in constant time, with regard to simulation length, for any prediction. We find that a single UCMax model, trained only on randomized materials, generalizes well to a suite of challenging multi-disciplinary inverse problems, showing strong efficacy and speed improvements up to 96\% on multi-angle optical tomographic reconstruction, beam shaping, and freeform inverse design of photonic devices, with no further training. UCMax does not require any separate, problem-specific adjoint model \cite{Jiang2019} or external method to compute design gradients, producing them directly through backpropagation. Our findings demonstrate the efficacy and feasibility of a single-model, verifiable neural surrogate to existing scattering simulators, and our conditioning and training methods are directly applicable to any ODE, PDE, or system thereof with an existing time-domain iterative solver, a class which includes the Navier-Stokes equations\cite{nav}, diffusion\cite{Um20103DTS}, and the Schr\"odinger equation \cite{Wang2010}.

\section{Results}
\subsection{Conditioning and training the unified neural simulator}
The UCMax architecture (diagrammed in Fig. 1) consists of an attentional\cite{Vaswani2017}, convolutional U-Net\cite{Ronneberger2015} which maps from a given voxel-based configuration of RIs to complex-valued field predictions for $E_z$, $H_x$, and $H_y$. In order to allow variable simulation conditions, we condition the inputs of each layer with pre-encoded wavelength, illumination, and time information. Directly aggregating positional encodings of each of these values via summation leads to ambiguity, however, as any combination of input values would produce the same value regardless of permutation. As such, we first individually positionally encode each simulation condition using sinusoidal encodings \cite{Vaswani2017}, process each with a separate small, multi-layer perceptron (MLP), then aggregate via summation. Then, this encoding vector is processed by a set of small, layer-specific MLPs, producing encoded vectors with the same size as the channel dimension of each layer. Finally, each encoding vector is added, along the channel dimension, to the inputs of its respective layer, prior to the attentional calculation. Compared with concurrent work on conditioning \cite{lupoiu2024}, UCMax requires only a single network pass per inference, functions for near-arbitrary materials, and exhibits excellent accuracy when conditioning with all conditioning variables (wavelength, angle, timestep) simultaneously. 

When training neural EM wave simulators in the context of specific inverse problems, it is common to assemble a focused dataset of training devices, simulated with existing finite-difference methods, which samples a small, targeted subset of the design space. When designing a photonic device with a specific figure of merit, for instance, a set of high-performing devices can be designed with conventional means, then a model supervised with this dataset to generate or optimize similar devices. As we wish for the UCMax model to be universally applicable to all inverse scattering tasks, we instead opt for an online, uniform sampling approach randomly selecting wavelength $\lambda$, plane wave input angle $\theta$, timestep $t$, and the RIs at each of $D$ voxels $\{n_1,...,n_D\}$ uniformly. Each random sample is used exactly once during training, then discarded, and a set of new random samples is selected for each epoch. 

Note that $D$ indicates the number of variable voxels, equivalent to the number of DoFs, and $N$ indicates the total number of voxels. In the following demonstrations, we consider $D=2560$ and $N=4096$, with the variable DoFs arranged into a $64 \times 40$ strip surrounded by air (RI of 1) to the left and right, forming a $64 \times 64$ region. This embedded strip is illuminated from the left by a configurable plane wave. Further training and system configuration specifics are provided in Methods.

We find that the UCMax model can be readily fit with low error to randomly selected, non-repeated data during training. Because this data is an unbiased stochastic selection from the ($D+3$)-dimensional hypercube of all possible systems (each dimension being a DoF or simulation parameter, and all but the timestep being continuous) this indicates an ability to accurately simulate any given possible system with high statistical probability. Described in Sections 2.5-7, we find that the so-trained model addresses well a broad range of structured and unstructured systems.
 
The applicability of UCMax is further extended in several ways. Firstly, due to optical linearity, any input wavefront can be readily represented as a coherent superposition of scaled, phase shifted plane waves. As UCMax predicts the full complex fields for arbitrary plane waves, a set of plane waves can be computed in parallel and assembled to simulate any illumination pattern. Broadband illumination can be handled similarly, by superimposing single-wavelength predictions. Similarly, one can simulate illumination from the reverse or sides of a material by mirroring or rotating the material, respectively, before NN inference. While any given UCMax model has a maximum addressable system size (the height and width of the randomized training data), this maximum system size can be defined arbitrarily at training time.

\begin{figure}
    \centering
    \includegraphics[width=\textwidth]{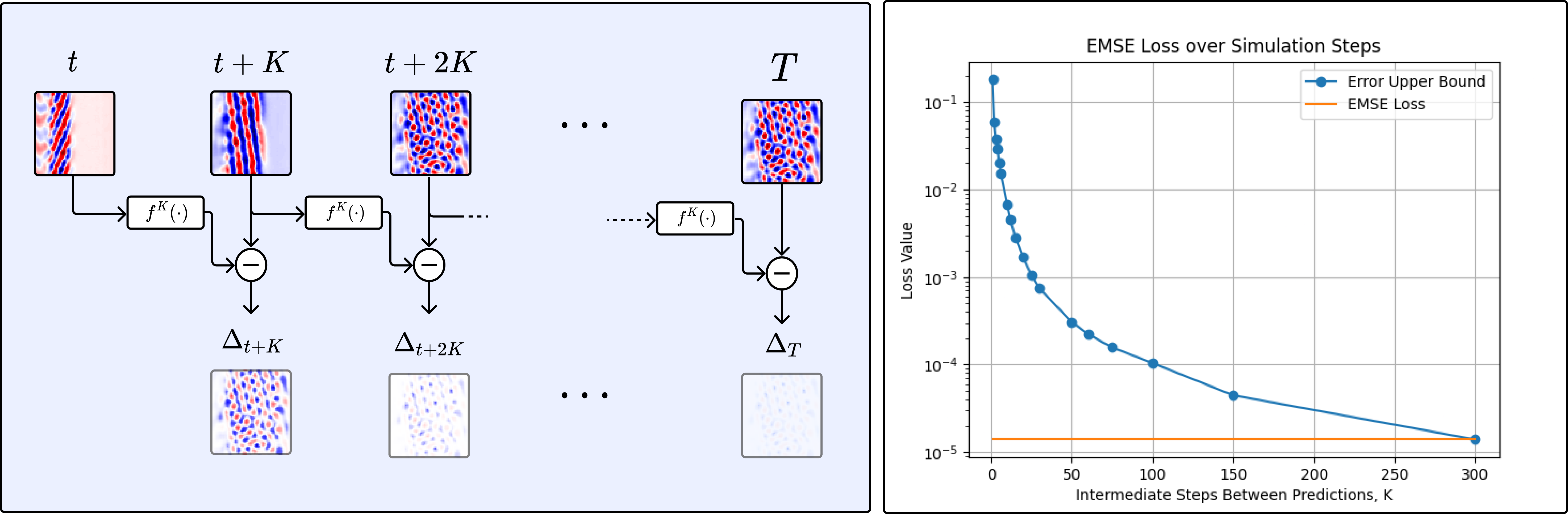}
    \caption{(left) Diagram of the error bound calculation method. (right) The error upper bound (in blue) and true error (in orange) at $T=300$ for different values of $K$. Note that the bound is asymptotic to the true error as $K$ approaches $T$.}
    \label{fig:enter-label}
\end{figure}
\subsection{Direct prediction of intermediate temporal states}
The most common approach for training a NN surrogate of an EM wave solver is direct prediction of final states, in which any intermediate states produced by the underlying simulator are ignored, and the NN is supervised to directly predict only the final converged state of the simulator. This approach is commonly fast compared to the conventional simulator, as it requires only a single step for inference. For time-stepping methods like FDTD, however, a great deal of intermediate, physically relevant information is necessarily produced in each simulation, namely the electromagnetic states at non-final timesteps, which existing direct prediction methods discard. In contrast to direct prediction, a recurrent approach \cite{hughes2019} which iteratively steps through each timestep of the simulation can in principle make use of these intermediate steps as training data. Such a recurrent approach, however, requires a sequence of $T$ steps for a $T$-timestep simulation and is in fact reducible to the iterative application of the time-domain update equations used in FDTD. As such, prediction of and training on intermediate states with existing methods is not possible without compromising inference speed.

UCMax is able to train on, and predict, intermediate temporal states generated by the underlying simulator, in $O(1)$ time for all such states. To achieve this, we condition the NN with a pre-encoded vector indicating the desired timestep, a mechanism similar to those used in diffusion-based image generation algorithms. During training, the network is trained with a mix of intermediate and final timesteps from the current batch of training simulations. In addition to making these intermediate states predictable at inference time in $O(1)$ time regardless of the timestep, training on these intermediate states has a regularizing effect, significantly improving the model's generalization ability as measured by the mean squared error (MSE) of the final timestep, as shown in Table 1. This effect is likely due to the model's exposure via this method to all of the intermediate physical dynamics, in contrast to simply training on the final state. The optical propagation process is self-similar in the sense that the EM fields at any adjacent timesteps are related by the FDTD update kernel, which is fixed for any given simulation, and therefore the task of predicting the EM fields at a given timestep has a close relationship with prediction for all other timesteps. By supervising on and being able to predict this entire temporal propagation process, the model is guided toward a more physically realistic mapping between RIs and EM fields, without the need for additional ground truth simulations. Additional training details are provided in Methods.

\subsection{Error upper bounds using direct prediction}
A significant issue in the use of NN-based surrogates for partial differential equation (PDE) simulation is in efficiently providing guarantees, at inference time, that the error of a prediction, when compared with the ground truth, is below a certain threshold. Residual-based physics-informed methods\cite{Lim2022} are often used to meet this need, but such residuals lack a quantitative correspondence with prediction error. Indeed, systems with the same supervised error can exhibit widely disparate physics-informed residuals, based on simulation specifics.

Using UCMax's ability to predict intermediate simulation states in $O(1)$ time, one can efficiently compute a rigorous upper bound on the prediction error for any prediction. This bound functions for any system without optical gain. For a $T$-step simulation with user-configurable positive integer $K\leq T$, this error bound can be computed in $O(K)$ time and $O(T/K)$ memory, where generally $K<<T$ in practice. As shown in Fig. 2, the tightness of the upper bound is configurable by $K$, recovering an exact prediction error when $K=T$. 

As diagrammed in Fig. 2, the error bound is calculated by querying the NN in parallel for EM field predictions at every $K$ timesteps $\{Y_K,Y_{2K},...,Y_{T-K},Y_T \}$, where we assume the initial condition $Y_0$ is known (and commonly $Y_0 = \mathbf{0}$). We then apply $K$ steps of the ground truth FDTD algorithm $f^K(\cdot)$ in parallel to the initial condition and all but the final prediction, producing $\{f^K(Y_0),f^K(Y_{K}),...,f^K(Y_{T-K}) \}$ . We now subtract the two sets, defining $\{\Delta_K,...,\Delta_T\} := \{Y_K-f^K(Y_0),...,Y_T-f^K(Y_{T-K})\}$ .  We consider the energy mean squared error across a simulation instance with $N$ total voxels, 
$$EMSE(Y_T,\hat{Y}_T) = \frac{1}{N}\sum_{i=1}^N n_i^2(Y_{T,i}-\hat{Y}_{T,i})$$
a variant of mean squared error which incorporates the RIs of a system and thereby calculates the optical energy in an error field. It can be proved that
$$EMSE(Y_T,\hat{Y}_T) \leq \frac{T}{N} \sum_{t=0}^T \sum_{i=1}^N (\Delta_{t,i}n_i)^2$$
A full derivation of this bound's validity is provided in Methods.

\subsection{Temporal and reversed-temporal physics-informed loss functions}
Because UCMax predicts the intermediate states of transient simulations, existing physics-informed training methods based on the wave equation \cite{Lim2022} are not applicable, as these only function for steady state solutions. Despite this, and in addition to the physical regularization provided by direct prediction, UCMax admits a family of efficiently-calculated physics-informed loss functions which provide further improvements to model generalization. To compute these loss functions, we apply $M$ steps of the FDTD simulator to the NN's batch of EM field prediction during training, producing $\{f^M(Y_{t_1}),f^M(Y_{t_2}),...,f^M(Y_{t_{batch}}) \}$ . Each step is then compared to the ground truth EM field at that respective time, yielding, including the supervised loss, 
$$\mathcal{L}_{phys,fwd} = \sum_{j=0}^M MSE(\hat{Y}_{t+j}, f^{j}(Y_t))$$

This loss is then backpropagated, with gradients flowing back through the simulator steps for the physics-informed losses.

We further implemented a time-reversed physics-informed loss which takes advantage of the reversibility of Maxwell's equations. Given a time-reversed FDTD simulator $f^{-j}(Y_t)=Y_{t-j} $, we can now also take M steps in the time-reversed direction, producing $\{f^{-M}(\hat{Y}_{t_1}),f^{-M}(\hat{Y}_{t_2}),...,f^{-M}(\hat{Y}_{t_{batch}}) \}$  and yielding the full loss function
$$\mathcal{L}_{phys,full} = \sum_{j=-M}^M MSE(\hat{Y}_{t+j}, f^{j}(Y_t))$$

These loss functions can be calculated in parallel for any batch size, require $O(M)$ time for both the forward and combined forward-reversed cases, and function both for transient and steady-state cases. A comparison of trainings with various loss functions and training approaches is shown in Table 1.

\begin{figure}
    \centering
    \includegraphics[width=0.80\textwidth]{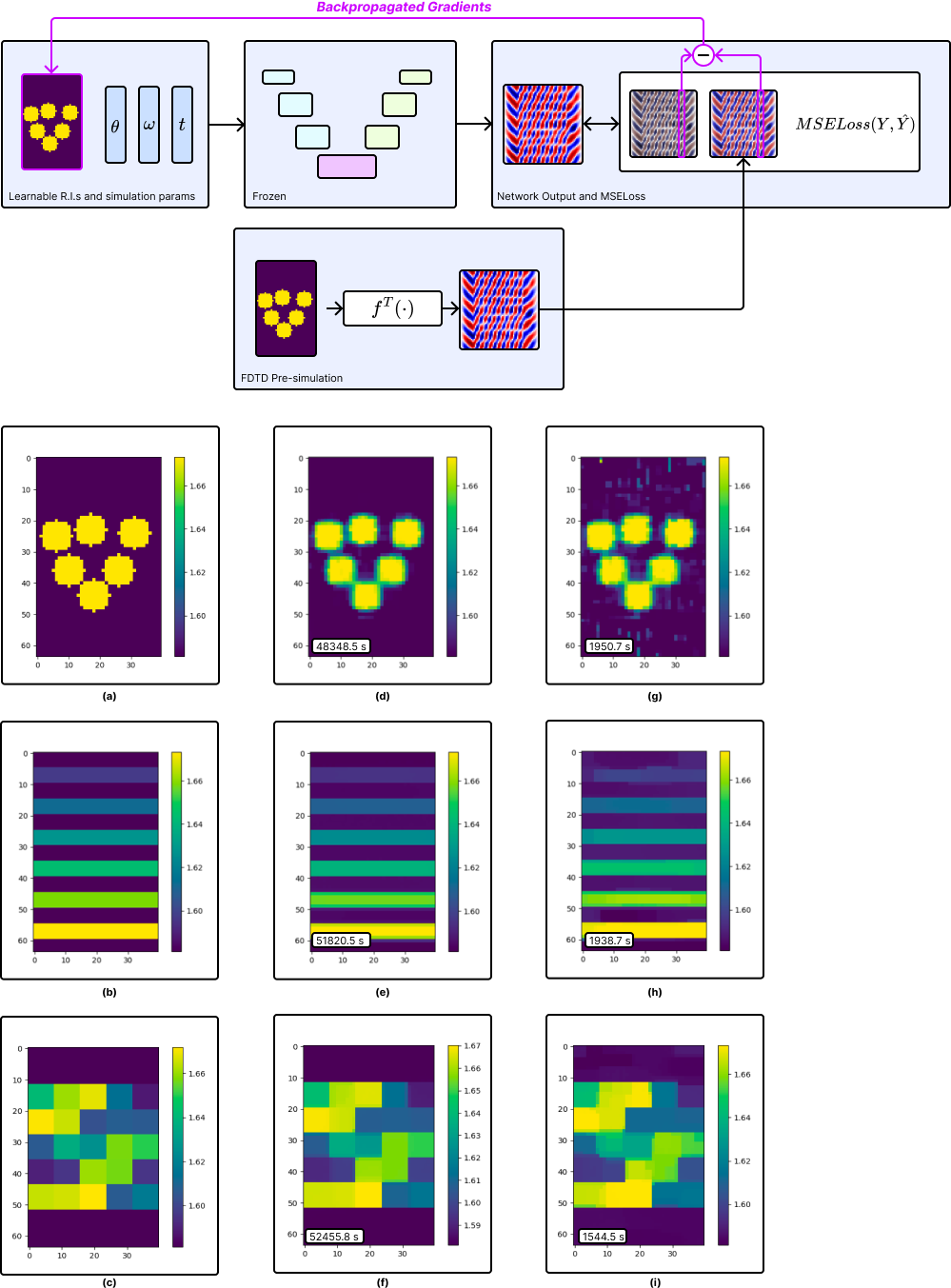}
    \caption{(top) Diagram of tomographic reconstruction. An initialization is iteratively updated based on tomographic readings from a ground truth simulation. (a-c) Target refractive indices for microspheres, multi-material nanopillars, and a randomized multi-material mosaic. (d-f) FDTD reconstructed refractive indices. (g-i) UCMax reconstructed refractive indices.}
    \label{fig:enter-label}
\end{figure}
\subsection{Tomographic reconstruction of arbitrary materials}
In the field of computational imaging, many problems take the form of a reconstructive process \cite{Hansen2015,Okawa2023,Yoon2015,Park2015}, requiring a series of differentiable electromagnetic wave simulations. These problems are often time-sensitive, and as such using physically accurate full-wave models is rarely feasible. Instead, reduced order models such as first-Born \cite{Park2018}, or multi-Born \cite{Chen2020} are used, which simplify the scattering problem to a driving field interacting with one or more scattering potentials. This works for thin, low-scattering materials but fails for systems with substantial multiple scattering. Addressing computational imaging problems with neural network-based simulators is a compelling prospect, as both high speed and high accuracy may be achievable. Forming such a neural simulator for computational imaging has proved challenging, however, as the structure of a sample is rarely known beforehand and may lack rigid structure, making parameterization with a small number of DoFs infeasible.

To demonstrate UCMax as a surrogate model for freeform computational imaging problems, we consider the case of optical tomography (OT). OT involves illuminating a volumetric scattering material with a succession of angled plane waves, recording images or complex-valued wavefronts from each experiment, then using a computer model of the illumination-material system to iteratively reconstruct a quantitative RI-based representation of the illuminated material. This approach requires no artificial staining or fixing, allowing the imaging of live, mobile samples, something of particular utility for oncology and embryo development. Like other computational imaging methods, however, the accuracy of OT is limited by the quality of the model. 

Diagrammed in Fig. 3(a) and with results shown in Fig. 3(b-j), we consider OT reconstruction of a cluster of microspheres, a multiple-material nanopillar device, and a mosaic of randomized materials. Using FDTD as a ground truth simulator, each is illuminated from the left with a 450 nm plane wave at a series of 100 input angles, evenly spaced between -80 and 80 degrees, and the wavefront is sampled immediately to the right of the system. For the microspheres and mosaic, we additionally simulate 360 degree illumination by rotating the 40 by 40 voxel square in the center of the system is by 90, 180, and 270 degrees, then illuminating with the same set of input angles, for a total of 400 illuminations.

These target wavefronts are then used in the reconstructive process. In the reconstructive process, an initial RI guess of 1.5 for the rectangular region is used, and the surrogate solver is used to differentiably simulate this guess for each input angle, including the rotated systems for the microspheres and mosaic. The batch of predicted wavefronts is then compared to the target wavefronts with mean squared error (MSE), and this error is backpropagated through the model to update the reconstruction. We additionally apply a total variation regularizer to smooth reconstruction dynamics.  This process is repeated until convergence (in this case 5000 times). As diagrammed in Fig. 3(a) and with results shown in Fig. 3(b-j), we compare using UCMax versus differentiable FDTD as the reconstruction model, with the MSE of the final reconstruction with the ground truth device inset, along with the simulation time. We find that the model produces reconstructions nearly indistinguishable from the FDTD-based reconstruction, while reducing reconstruction times by up to 96\%.
\begin{figure}
    \centering
    \includegraphics[width=0.75\textwidth]{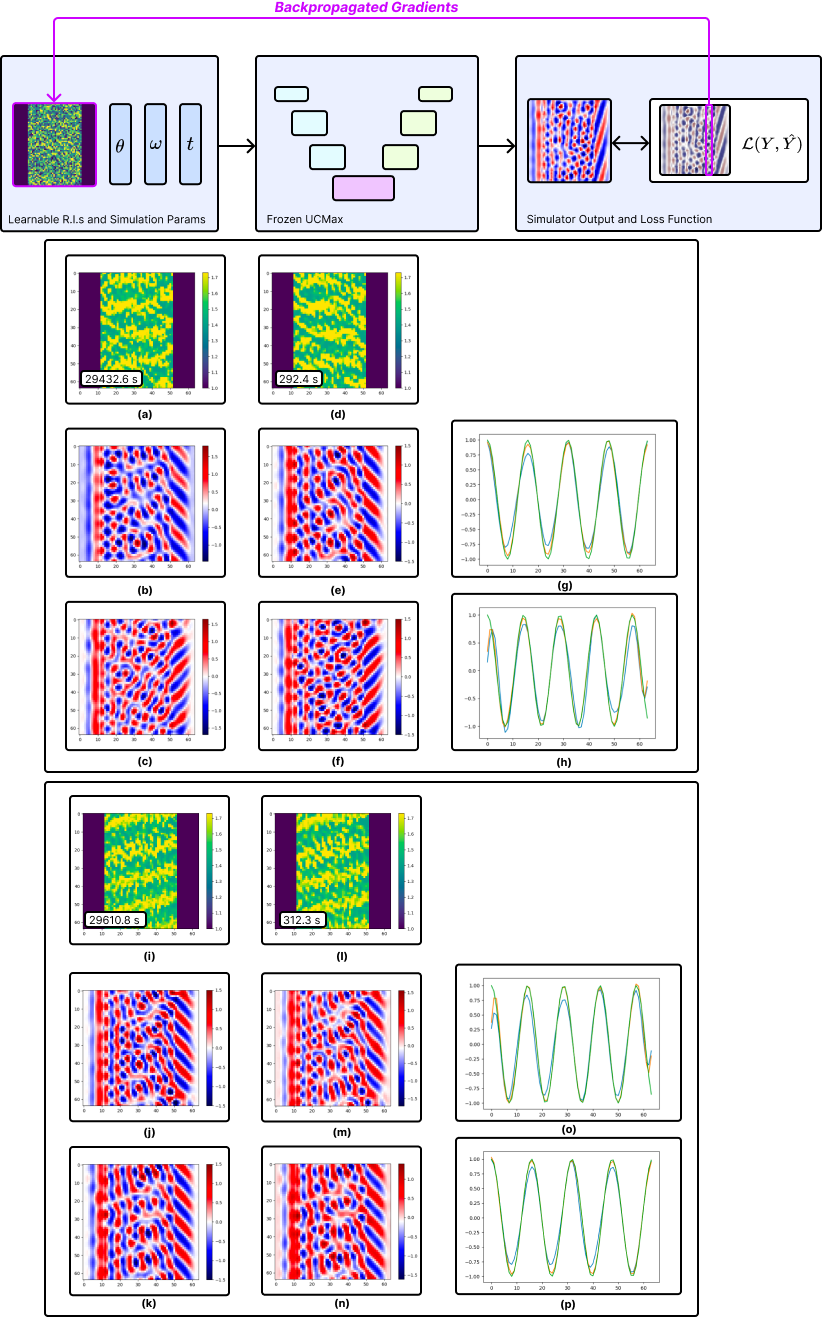}
    \caption{(top) Diagram of the inverse design process. (a-c) Refractive indices for the FDTD-designed spectral splitter, EM fields of the light with $\lambda=450$nm coupling to +80 degrees, and EM fields of light with $\lambda = 500$nm coupling to -80 degrees, respectively. (d-f) The same images for a UCMax-designed splitter. (i-k) Results for an FDTD-designed multi-wavelength coupler, with $\lambda=450$nm and $\lambda=500$nm both coupled to +80 degree outputs (j-m) The same results for a UCMax-designed coupler. Graphs (g), (h), (o), and (p) compare the electromagnetic field amplitudes along a vertical strip one voxel to the right of the designed devices, showing target field (green), FDTD-designed field (orange), and UCMax-designed field (blue).}
    \label{fig:enter-label}
\end{figure}
\subsection{Inverse design of freeform photonic devices}
Unlike traditional photonic design, in which devices are hand-crafted as compositions of existing components, photonic inverse design \cite{Molesky2018} allows the flexible computational optimization of devices based on chosen figures of merit (FoMs). This approach has been used to design devices such as metamaterials\cite{Yu2014}, beam splitters\cite{Piggott2015}, and couplers \cite{Su2018} of previously unrealizable flexibility and compactness. This process requires a series of differentiable simulations, the speed and scaling of which practically limit the size and complexity of devices which can be designed. Composite FoMs which require multiple simulations to compute a single update, such as for a device mapping multiple individual inputs to multiple individual outputs, are especially limited by this. Devices must often be simplified to ameliorate these issues. 

UCMax can be used as a surrogate model for the design of freeform photonic devices. Diagrammed at the top of Fig. 4 and with results shown in Fig. 4(a-p), we consider as a demonstration the problem of designing volumetric, freeform spectrally selective splitters and couplers, in which a plane wave incident on the left of the device is coupled to rightward-propagating output waves based on its wavelength. In the first example, we design a splitter which maps a normally incident plane wave of 450 nm light to an outgoing plane wave of an angle 80 degrees from normal, and a 500 nm normal plane wave to -80 degrees from normal. In the second example, we design a coupler in which both 450 nm and 500 nm normally incident plane waves are mapped to 80 degree plane wave outputs. For each of these inverse design problems, compared to designs based on FDTD, the NN based designs produce commensurate performance with approximately 90\% faster runtime. 
\begin{figure}
    \centering
    \includegraphics[width=0.90\textwidth]{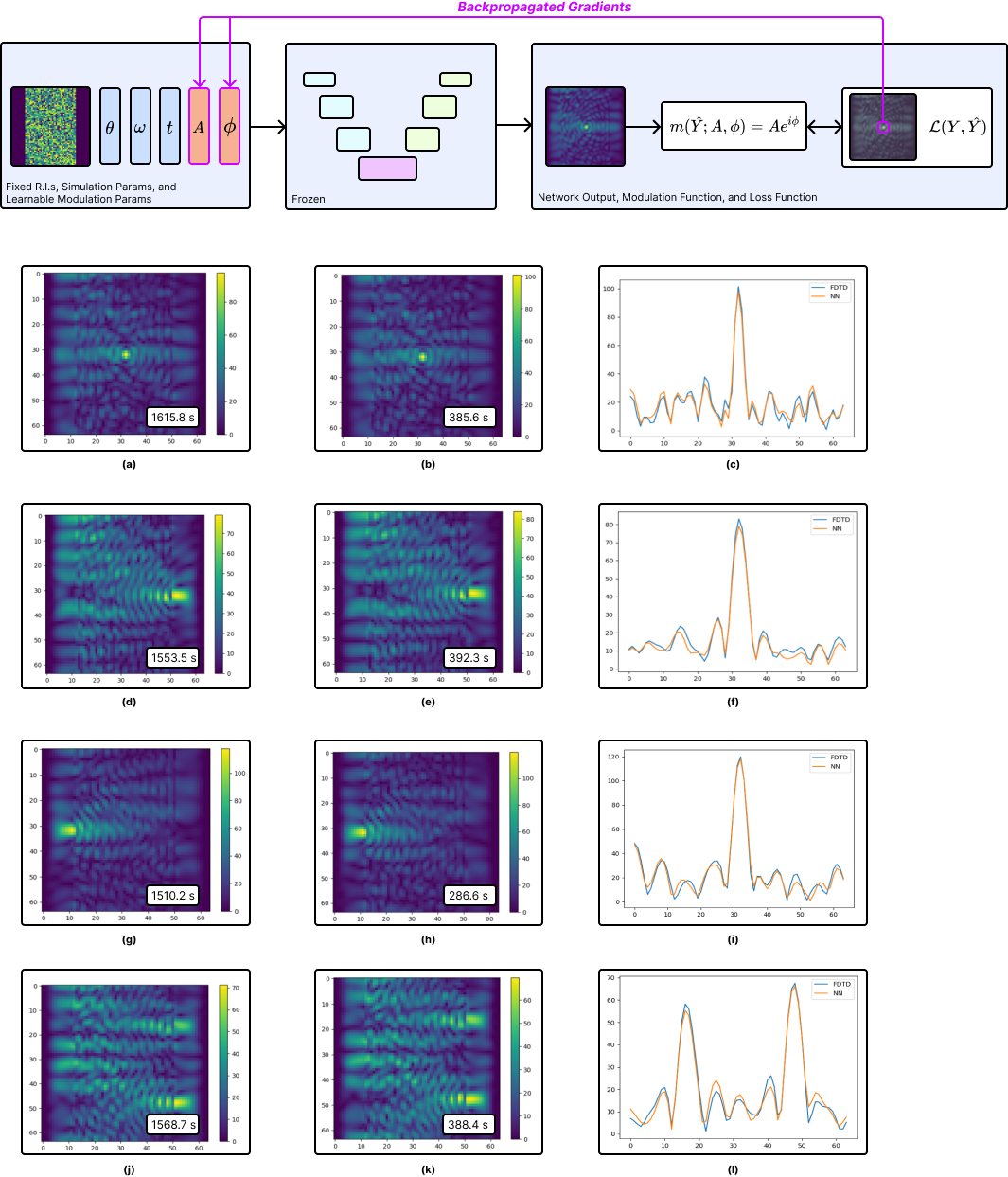}
    \caption{(top) Diagram of the wavefront shaping process. Note that the amplitudes $A$ and the phases $\phi$ are the only parameters being learned. (a-c), (d-f), (g-h), and (j-k) show the average power of the electromagnetic fields for an FDTD-designed source, UCMax-designed source, and a graph that compares optical power at a vertical strip positioned at the chosen focal location, respectively. Runtimes are noted for each design.}
    \label{fig:enter-label}
\end{figure}
\subsection{Wavefront shaping through high-contrast disordered media}
Imaging deep into highly scattering biological tissue is a topic of foundational importance in medicine and biology \cite{Yoon2022,Helmchen2005,Feuchtinger2016,Gigan2022}. To achieve this, it is necessary to bring light to a tight focus deep within tissue, which is a challenge due to the optical scattering introduced by the intervening tissue layers. Additionally, in many deep tissue imaging applications, especially in vivo, placing reference guidestars \cite{Horstmeyer2015} behind the scattering material to allow direct calibration is not feasible. In such cases, a partial model of the tissue may be formed, for instance with reflection-mode OT, and the illumination tuned in order to create concentrated light in or behind the material. The success of this process relies heavily on the quality of the model, as any modeling error builds up as the light penetrates deeper into the tissue. This problem is also time-sensitive, as biological tissues change over time due to desiccation, movement or deformation.

Here, we demonstrate UCMax on tasks involving a high RI-contrast randomized scattering medium. We consider the case of focusing to a single point within the medium, a single point past the medium, multiple points past the medium, and a single point reflected from the medium. We consider a controlled illumination which is a weighted, phase-shifted coherent superposition of a set of 100 plane waves at angles evenly spaced from -80 to 80 degrees. For these tasks, the material is fixed, and amplitude weights $a_i \in  [ 0,1 ]$ and phase shifts $\phi_i \in [0,2\pi]$  are learned for each plane wave $i$. For the in-material task, we consider the vertical plane in the center of the scattering material, maximizing the average power at the center of that plane. Similarly, for the through-material task, we maximize the average power at the center of a plane positioned ten voxels to the right of the scattering material. For the two-point task, we do the same but with two points each 16 voxels equidistant from the center, vertically. Finally, for the reflection-mode task, we do the same but at the vertical plane one voxel to the left of the material. For all tasks, the illumination parameters are iteratively updated with gradients computed by backpropagation through the model, over 1000 steps, and the final parameters are evaluated with FDTD. This is compared with using FDTD as the model for the illumination-updating process. The results of this are shown in Fig. 5, with near-identical outputs produced by each method and UCMax providing speed improvements of up to 80\%.

\section{Discussion}

Neural network-based simulators offer the promise of EM wave simulations which simultaneously achieve near-real-time speeds and uncompromising accuracy. Existing approaches, however, only scale to simple, structured problems with approximately 20 DoFs, only address narrowly tailored problem classes, and lack a quantitative method for verifying the accuracy of predictions. In this work, we demonstrate that an attentional multi-conditioning technique, with a corresponding time-domain verification scheme and randomized training method, is able to produce a model capable of arbitrary, problem-agnostic simulations within a set of broad, configurable spatial and parametric bounds, with efficiently verifiable inference-time accuracy. Demonstrated across a suite of multi-disciplinary inverse problems, we find that, compared to the ground truth FDTD simulator, our UCMax model reliably achieves 80-96\% faster operation while providing commensurate reconstruction and design quality. 

Our findings indicate a practical path toward universally applicable, verifiably accurate neural surrogates for existing scattering simulators. We believe that foundation models trained using the techniques developed here will allow more practitioners to realize the benefits of NN-based simulation--without the requirement of individualized datasets or extensive machine learning expertise. 

During training, UCMax fits to uniformly randomly selected refractive index configurations and simulation conditions. As such, the trained UCMax model has the ability to, with high statistical probability, accurately simulate any given set of RIs and conditions. This does not, however, rule out the possibility of adversarial examples and edge cases for which the model performs poorly. Addressing these edge cases and ensuring globally robust prediction accuracy may be the subject of future work.

We expect the methods demonstrated in this work to be straightforwardly generalizable to other physics simulation tasks. NN-based simulators of fluid dynamics, for instance, exhibit similar limitations to the existing NN-based EM wave simulators discussed here and would likely benefit from the training, conditioning, and intermediate timestep prediction capabilities here developed. Similarly, the time-domain verification methods developed here are equally applicable to any linear, energy-conserving or energy-dissipating PDE or ODE. For instance, these methods could be used to verify predictions for NN-based simulators of heat conduction. 

\section{Methods}

\subsection{Training details}
For demonstration purposes, we choose a continuous wavelength range between red (450 nm) and blue (750 nm), a continuous angular spectrum of input plane waves between -80 and 80 degrees, timesteps from 0 to 300, a RI range between 1 (air) and 2, and number of DoFs $D=2560$, as these ranges are well addressed by the underlying FDTD simulator. We arrange the 2560 DoFs into a $64 \times 40$ region, padded with constant-valued RIs of 1 to form a $64\times 64$ region with total number of voxels $N = 4096$, with a configurable additive plane wave source illuminating the material from the left, periodic boundry conditions on the top and bottom, and perfectly matched layer (PML) absorbing boundary conditions on the left and right. Each voxel is taken to be a square with a side length of 45 nm. Before the region of RI values is provided to the NN, these values are augmented with the location of the source, with a vertical strip one voxel to the left of the material being set to a gradient of negative values ranging from -1 at the top of the source to -2 at the bottom. This is done so that the NN has a consistent point of reference available. A combined temporal and reverse-temporal physics-informed loss function was used, as described in 2.4, with M=1. The UCMax model was trained, using randomized sampling from all parameter ranges, for 100,000 epochs on a single A6000 GPU over a period of 8 days, with each epoch training from 50 ground truth FDTD simulations, achieving a final training MSE of 0.001. For the ground truth algorithm, a purpose-built, GPU-based, differentiable FDTD simulator is used, written in Python 3 using the PyTorch framework. A comparison of training approaches was performed, with each approach being trained for 2,000 epochs, then tested on a dataset of 200 randomized materials, with the RI distribution from which voxel is drawn being $\frac{1}{\mathcal{U}(1/3,1)}$ as compared to the $\mathcal{U}(1,3)$ used during training. As the new distribution is likely to produce significantly different materials from those used during training, the NN must generalize well to perform well on these materials. Results are provided in Table 1, with the use of intermediate temporal states and temporal physics-informed loss functions providing better generalization than standard training based on direct prediction.

\begin{table}[]
    \centering
    \begin{tabular}{|p{0.55\linewidth} | p{0.35\linewidth}|}
        \hline
        \centering
        Training Configuration & Out of Distribution Dataset MSE\\
        \hline
        \hline
        \centering
        Only final states, only MSE loss & 0.895 \\
        \hline
        \centering
        Both intermediate and final states, only MSE loss & 0.721 \\
        \hline
        \centering
        Both intermediate and final states, temporal physics-informed loss, $M=1$ & 0.719 \\
        \hline
        \centering
        Both intermediate and final states, temporal and reverse-temporal physics-informed loss, $M=1$ & 0.649 \\
        \hline

    \end{tabular}
    \caption{Performance of training approaches after 2000 training epochs, measured on a 200-example out-of-distribution dataset requiring generalization.}
    \label{tab:my_label}
\end{table}

\subsection{Proof of error upper bound}

The operating principle of the error upper bound is diagrammed in Fig. 2. For a given set of inputs, the NN is queried in a single batch for every K timesteps, producing field predictions $\{\hat{Y}_{t0}, ...,\hat{Y}_{T-K}, \hat{Y}_T\}$, where we here consider only the predicted E field state and represent it as a 1-dimensional vector for ease of exposition, without loss of generality. To all but the final state at $T$, in parallel, we apply K steps of the ground truth FDTD algorithm $f^K(\cdot)$, producing $\{f^K(\hat{Y}_{t0}), ..., f^K(\hat{Y}_{T-K})\}$. 

In the following proof of this error bound, we consider the energy mean squared error (EMSE) as our loss function:

$$EMSE(Y_t,\hat{Y}_t) =\frac{1}{N} \sum_i^N n_i^2(Y_{t,i}-\hat{Y}_{t,i})^2 = \frac{1}{N} \sum_i^N n_i^2(\epsilon_{t,i})^2$$

We consider an error bound computation for single steps, $f(\cdot)$. If multiple steps are desired (a tighter bound requiring more compute), one can simply replace $f(\cdot)$ with $f^K(\cdot)$. We derive this error bound for the final state at the final timestep $T$, without loss of generality. Note that, for notational convenience, the influence of the source is implicitly handled by $f(\cdot)$ such that, for a sum of multiple $f(\cdot)$, the source is only added once. 

We first begin by defining the relationship between FDTD generated EM fields, denoted $Y_t$, and neural network generated EM fields $\hat{Y}_t$ at time $t$. We can model each neural network prediction $\hat{Y}_t$ as a linear combination of a ground truth simulation $f^t(Y_0)$ and an accumulation of perturbations, each defined by $\Delta_{t} = Y_{t} - f(Y_{t-1})$. As a result, we can define each network prediction iteratively:

$$\hat{Y_0} = Y_0 = \mathbf{0}$$
$$\hat{Y}_1 = f(Y_0) + \Delta_1$$
$$\hat{Y}_2 = f^2(Y_0) + f^1(\Delta_1) + \Delta_2$$
$$\hat{Y}_3 = f^3(Y_0) + f^2(\Delta_1) + f(\Delta_2) + \Delta_3$$
$$\cdots$$
$$\hat{Y}_T = f^T(Y_0) + f^{T-1}(\Delta_1) + f^{T-2}(\Delta_2) + \cdots + \Delta_T$$
Grouping the error terms together, we note that this is equivalent to the error field at timestep $T$, $\epsilon_T$.
$$\sum_{t=0}^T f^{T-t}(\Delta_{t}) = \epsilon_T$$
We then apply the Cauchy-Schwartz inequality, written as: $$\frac{(\sum_{n=0}^N u_n)^2}{\sum_{n=0}^N v_n} \leq \sum_{n=0}^N \frac{u_n^2}{v_n}$$
Defining $\mathbf{v} = \mathbf{1}$, and $\mathbf{u} = \begin{bmatrix} 0 & f^{T-1}(\Delta_{T-1}) & f^{T-2}(\Delta_{T-2})& \cdots & \Delta_{T} \end{bmatrix}$ yields:
$$\frac{(\sum_{t=1}^T f^{T-t}(\Delta_{t}))^2}{T} \leq \sum_{t=1}^T f^{T-t}(\Delta_{t})^2$$
We can then substitute in our expression for $\epsilon$:
$$(\sum_{t=1}^T f^{T-t}(\Delta_{t}))^2 \leq T\sum_{t=1}^T f^{T-t}(\Delta_{t})^2$$
$$\epsilon^2 \leq T\sum_{t=1}^T f^{T-t}(\Delta_{t})^2$$
Multiplying by the square of the refractive index and summing over spatial dimensions:
$$\sum_{i=1}^Nn_i
^2\epsilon_i^2 \leq T\sum_{i=1}^N n_i^2\sum_{t=1}^T f^{T-t}(\Delta_{t,i})^2$$
$$\sum_{i=1}^N n_i^2\epsilon_i^2 \leq T\sum_{t=1}^T \sum_{i=1}^N( n_if^{T-t}(\Delta_{t,i}))^2$$
Note that we can define the total energy within each time-advanced $ f^{T-t}(\Delta_{t})$ as:
$$\mathcal{U}_t = \sum_{i=1}^N( n_if^{T-t}(\Delta_{t,i}))^2$$
Due to energy conservation, for any system without gain:
$$\mathcal{U}_t \leq \sum_{i=1}^N( n_i\Delta_{t,i})^2$$
Thus:
$$\frac{1}{N}\sum_{i=1}^Nn_i^2\epsilon_i^2 \leq \frac{T}{N}\sum_{t=1}^T\sum_{i=1}^N( n_i\Delta_{t,i})^2$$
Finally, by the definition of EMSE:
$$EMSE(Y_T,\hat{Y}_T)\leq \frac{T}{N}\sum_{t=1}^T\sum_{i=1}^N( n_i\Delta_{t,i})^2$$

\bibliography{export2}










\end{document}